\documentclass[lettersize,journal]{IEEEtran}
    
\usepackage{caption}
\usepackage{multirow}
\usepackage{amsmath,amsfonts}
\usepackage{algorithmic}
\usepackage{algorithm}
\usepackage{array}
\usepackage[caption=false,font=normalsize,labelfont=sf,textfont=sf]{subfig}
\usepackage{textcomp}
\usepackage{stfloats}
\usepackage{url}
\usepackage{verbatim}
\usepackage{graphicx}
\usepackage{cite}
\usepackage[dvipsnames]{xcolor}
\renewcommand{\Vec}[1]{\mathbf{#1}}
\newcommand{\Mat}[1]{\mathbf{#1}}

\newcommand{\normtwo}[1]{\left \lVert#1 \right \rVert_2^2}
\newcommand{\norm}[1]{\left \lVert#1 \right \rVert}

\begin{document}

\title{Efficient Sound Field Reconstruction with Conditional Invertible Neural Networks}

\author{Xenofon Karakonstantis, \,
Efren Fernandez-Grande, \,
Peter Gerstoft
}





\maketitle

\begin{abstract}
\today \\
In this study, we introduce a method for estimating sound fields in reverberant environments using a conditional invertible neural network (CINN). Sound field reconstruction can be hindered by experimental errors, limited spatial data, model mismatches, and long inference times, leading to potentially flawed and prolonged characterizations. Further, the complexity of managing inherent uncertainties often escalates computational demands or is neglected in models. Our approach seeks to balance accuracy and computational efficiency, while incorporating uncertainty estimates to tailor reconstructions to specific needs. By training a CINN with Monte Carlo simulations of random wave fields, our method reduces the dependency on extensive datasets and enables inference from sparse experimental data. The CINN proves versatile at reconstructing Room Impulse Responses (RIRs), by acting either as a likelihood model for maximum a posteriori estimation or as an approximate posterior distribution through amortized Bayesian inference. Compared to traditional Bayesian methods, the CINN achieves similar accuracy with greater efficiency and without requiring its adaptation to distinct sound field conditions.
\end{abstract}



\section{Introduction}

\IEEEPARstart{W}{ith} spatial audio technologies promptly developing, there is an escalating need for sophisticated sound field reconstruction algorithms. This demand is driven by the need to replicate, analyze, and control auditory experiences. This surge of interest spans many fields, from architectural acoustics to audio signal processing, where researchers are driven to explore a wide range of computational methods and modeling techniques in pursuit of accurately represented sound fields. 

Sound field reconstruction involves recreating and characterizing sound field distributions within an enclosed space conditioned on limited pressure measurements. It finds extensive utility across multiple domains. Sound field analysis, facilitates the exploration of acoustic environments, providing insights into propagation characteristics, composite wave interactions, and sound decay within confined spaces \cite{Nolan2019,Berzborn2021}. In virtual reality, this methodology assists in crafting immersive auditory experiences, enhancing the realism of virtual environments \cite{Deppisch2023,Birnie2021a}. Moreover, applications in sound field control \cite{Heuchel2020, Abe2023},  personalized sound zones \cite{Vindrola2021, Cadavid2024} or reproduction methods \cite{Erdem2023,Zuo2020} benefit from prior knowledge of sound fields, enabling the creation of optimized auditory spaces to accommodate diverse preferences or needs within shared environments.
%
%

Traditional sound field reconstruction techniques often exploit spatial regularization methods to enforce smoothness or sparsity constraints on the reconstructed field. \cite{ Mignot2013, Verburg2018, Koyama2019, Antonello2017, Schmid2021} These methods aim to mitigate noise amplification using prior knowledge about the sound field's spatial composition and spectro-temporal properties. Additionally, spatial covariance-based techniques, such as kernel methods using Bessel functions, capture spatial interactions and propagate acoustic information across sensor arrays \cite{Ueno2018, CaviedesNozal2021, Ribeiro2023} Further advancements include exploring localized structures within the data via dictionary learning approaches \cite{Hahmann2021} or spatial decomposition models that incorporate the physics-based priors over the signal properties \cite{Pezzoli2022a, FernandezGrande2021}. However, these methods often show limited adaptability, requiring adjustments whenever the characteristics of a sound field change. They can also be time-consuming, making drawing inferences less efficient. Additionally, the analytical tools describing wave propagation usually focus on the frequency domain \cite{Williams1999}. In practical terms, this means that to achieve tasks such as sound field reproduction, it's necessary to use a separate model for each discrete frequency.

The emergence of deep learning, particularly the integration of neural networks, presents a promising avenue for significant advancement in this field. Deep learning can capture intricate patterns and statistical dependencies within sound field data \cite{Cobos2022, Goetz2023, Zea2023}, thereby enabling more accurate and efficient reconstruction processes. Recent years have seen a notable increase in deep learning within the sound field reconstruction literature \cite{Lluis2020, Pezzoli2022b, Karakonstantis2023, Miotello2024}. A pivotal aspect of deep learning for sound field reconstruction involves integrating principles of physics into the learning framework. The addition of physical constraints act as a form of inductive bias, guiding the learning process toward solutions that align with the underlying physical phenomena governing sound propagation. This incorporation can manifest explicitly by directly solving the equation describing wave propagation \cite{Shigemi2022, Karakonstantis2024}, or implicitly by constraining the deep learning model solutions within the range of feasible solutions \cite{Olivieri2021, Karakonstantis2023}.

%
%
Within the landscape of deep learning for parameter estimation, invertible neural networks (INNs) stand out based on the principles and applications of normalizing flows \cite{Papamakarios2021}. When conditioned on an auxiliary variable in the form of complementary data or measurements---thereby becoming CINNs---these networks have proven effective in inferring parameters across high-dimensional spaces \cite{Ardizzone2019,Padmanabha2021,Ardizzone20182022, Radev2022}. They can approximate complex probability distributions by employing several bijective transformations, an architecture that presents multiple advantages: It enables Bayesian inference, allows for efficient sampling, and facilitates the estimation of uncertainties. This combination of flexibility and computational efficiency enables real-time inference, making them very suitable for swiftly processing large-scale spatio-temporal data.

The principle that allows for this efficiency is the concept of ``amortization" over the computational expenses incurred during the initial phase of training the CINN. Amortized Bayesian inference efficiently delineates the association between observations and their posterior distributions using a conditional density estimator, notably a trained CINN \cite{Cranmer2020}. This approach amortizes the initial training cost, facilitating expedited inference for new observations using the pre-trained model. Consequently, the method capitalizes on the initial computational cost, enabling efficient posterior estimation for subsequent observations without retraining.

This study proposes a CINN architecture with a series of normalizing flows to address sound field reconstruction. We propose assimilating the forward process of plane wave propagation from random wave fields satisfying the Helmholtz equation so that the network is poised for broad applicability across acoustic environments. Unlike other methods, this methodological choice circumvents the complexities associated with explicit regularization and selection of specific models, as the network's training spans an extensive range of sound fields and frequencies. Moreover, the versatility of the proposed approach enables both accurate reconstruction of sound fields and a framework for rapid prediction of a sound field distribution, without compromising precision.

The bijective nature of CINNs also enables the model to not only expedite the sound field parameters but also to assess uncertainty, thereby providing a measure of confidence in its predictions. This is achieved by drawing samples from a known analytic base distribution and processing these through the CINN, thus deriving the parameter distribution post-inference. Consequently, our approach affords a robust measure of the predictive confidence, integral to sound field analysis or any desired post-processing application.

To validate the approach, we conduct experiments utilizing data acquired from a fully furnished auditorium and compare the results with a principled method for Bayesian inference, specifically a hierarchical plane wave model employing Markov Chain Monte Carlo (MCMC). Furthermore, we interpret the uncertainty provided in the invertible networks' predictions and evaluate its capabilities at RIR estimation.

The subsequent sections are structured as follows. Section \ref{sec:Methods} provides an overview of the methods employed, which includes the acoustic model, normalizing flows, details about the architecture and training of the CINN, and details on a hierarchical plane wave model baseline. Section \ref{sec:experimental_dataset} outlines the experimental dataset utilized for model validation, while Section \ref{sec:results} presents details specific to the training and inference procedures, along with the results and analysis. This includes discussions on a small-scale example, baseline comparisons, uncertainty quantification, and RIR reconstruction. Lastly, Section \ref{sec:conclusions} offers concluding remarks.
\section{Methods}\label{sec:Methods}
\subsection{Plane wave acoustic model}
\label{subsec:pwexpansion}
In an acoustic domain devoid of sources, the linear sound field is governed by the homogeneous Helmholtz equation, exhibiting time-harmonic dependence. This implies that at any position $\mathbf{r}_m = [x_m, y_m, z_m]^T$ the measured sound pressure within a spatially extended region $\Omega \in \mathbb{R}^3$ can be expressed as\cite{Williams1999}
\begin{equation}\label{eq:planewaveexpansion}
\mathbf{p} = \mathbf{H} \mathbf{x} + \mathbf{n},
\end{equation}
where $\mathbf{p}=\left[p_1,  \ldots, p_M\right]^T$ denotes the measured sound pressure vector. The dictionary $\mathbf{H} \in \mathbb{C}^{M \times N}$ comprises $N$ plane waves, with each element $\mathbf{H}_{mn}=\mathrm{e}^{j \mathbf{k}_n \cdot \mathbf{r}_m}$ reflecting the influence of the $n$-th plane wave on the $m$-th measurement position. The vector $\mathbf{x}=\left[x_1,  \ldots, x_N\right]^T$ contains the unknown plane wave coefficients, and $\mathbf{n}=\left[n_1, \ldots, n_M\right]^T$ represents the measurement noise at each observation point. The plane wave coefficients, associated with the direction $\mathbf{k}_n$, are crucial for the acoustic field's evaluation. The additive noise term $\mathbf{n}$ encompasses various sources of unwanted disturbances or measurement errors that may affect the accuracy of the measured sound pressure. Only propagating waves are included in this study, implying far-field assumptions.

The task of determining these plane wave coefficients is critical for interpolating or extrapolating the sound pressure at positions $\mathbf{r}_R=\left[x_R, y_R, z_R\right]^T$, where measurements are not available. This process is complicated by the inherent challenges of ill-posedness, rank deficiency, and underdetermination, primarily due to the discrepancy between the number of measured pressures $M$ and the number of plane waves $N$ used to reconstruct the sound field (i.e.  \(N>M\)).


%
\subsection{Normalizing flows}
\noindent
Normalizing flows are parameterized bijective transformations, designed to establish an inverse mapping between two distributions. Specifically, a function $\boldsymbol{g}$ and its exact inverse $\boldsymbol{g}^{-1}$, facilitate the conversion of $N$-dimensional latent variables $\Vec{z} \sim \pi_z(\Vec{z} )$ into $N$-dimensional representations of target data $\Vec{w} \sim \pi^{*}_w(\mathbf{w})$, and vice-versa such that
\begin{align}
    \boldsymbol{g}:& \, \Vec{z} \rightarrow \Vec{w}, \\
    \boldsymbol{g}^{-1}:& \, \Vec{w} \rightarrow \Vec{z}.
\end{align}
%

Given parameters $\boldsymbol{\theta}$ of the transformations $\boldsymbol{g}$ and $\boldsymbol{g}^{-1}$, the ultimate aim of a normalizing flow is to closely approximate a data distribution $\pi^{*}_w(\mathbf{w})$ through a probability density $q_{\theta}(\Vec{w})$ derived from the change-of-variable formula
\begin{align}\label{eq:change_of_var1}
q_{\theta}(\Vec{w}) 
&=\pi_z\left(\boldsymbol{g}^{-1}(\Vec{w}; \boldsymbol{\theta})\right)\left|\left(\frac{d \boldsymbol{g}^{-1}(\Vec{w}; \boldsymbol{\theta})}{d \Vec{w}}\right)\right|,
\end{align}
where $\left|\cdot\right|$ denotes the determinant of the Jacobian matrix of $\boldsymbol{g}^{-1}$, crucial for maintaining the computational efficiency of the parameterized transformation. An invertible neural network $\Vec{g}$ combines several normalizing flows that are parameterized by neural networks. This is done to achieve the mapping of a simple tractable distribution $\Vec{z}_0 \sim \mathcal{N}(\Vec{0}, \Vec{I})$ to the data space $\Vec{w}$ using $N_f$ such transformations. This compositional nature, structured as
\begin{equation}
\Vec{w} = \Vec{g}(\Vec{z}_0) = \boldsymbol{g}_{N_f} \circ \cdots \circ \boldsymbol{g}_0(\Vec{z}_0),
\end{equation}
ensures that the final variable $\Vec{w}$ is derived through an intricate yet computationally tractable process, effectively bridging simple analytical forms and the multifaceted nature of the data.

\subsection{CINNs for modeling plane wave propagation} \label{subsec:CINN_wave_model}
The Bayesian framework offers a structured approach for estimating the coefficients $\Vec{x}$ condition on the measured pressure $\Vec{p}$. These coefficients $\Vec{x}$ are estimated via the posterior probability distribution  $ \pi(\Vec{x}|\Vec{p}) $  derived from Bayes' theorem \cite{Rubin1995}
\begin{equation}\label{eq:bayes_theorem}
    \pi(\Vec{x}|\Vec{p}) = \frac{\pi(\Vec{p}|\Vec{x})\pi(\Vec{x})}{\pi(\Vec{p})},
\end{equation}
where $\pi(\Vec{p}|\Vec{x})$ represents the likelihood, $\pi(\Vec{x})$ is the prior and $\pi(\Vec{p})$ refers to the evidence. 

This research aims to develop a CINN that can closely approximate the posterior distribution $\pi(\Vec{x}\mid\Vec{p})$ over plane wave coefficients $\Vec{x}$. This involves an invertible function $\mathbf{g}: \mathbb{C}^N \rightarrow \mathbb{C}^N$, parameterized by $\boldsymbol{\theta}$ and conditioned on the pressure vector $\Vec{p}$. Following \cite{Ardizzone2019} and \cite{Radev2022} we define the approximation of the true posterior by
\begin{equation} \label{eq:CINN_approx_posterior}
    q_{\theta}(\Vec{x} \mid \Vec{p}) \approx \pi(\Vec{x} \mid \Vec{p})
\end{equation}
where a model aims to match the distribution of $\Vec{x}$ and $\Vec{p}$ across all possible values. The model reparameterizes $q_{\theta}$ using network $\Vec{g}$, to transform between coefficients $\Vec{x}$ and a complex Gaussian latent variable $\Vec{z}_0$ as follows
\begin{equation}
    \Vec{x} \sim q_{\theta}(\Vec{x} \mid \Vec{p}) \Longleftrightarrow \Vec{x}=\mathbf{g}(\mathbf{z}_0 ; \mathbf{p}, \boldsymbol{\theta}) \text{ with } \mathbf{z}_0 \sim \mathcal{N}_c(\mathbf{z}_0 \mid \mathbf{0}, \mathbf{I}).
\end{equation}

To ensure that the outputs of $\mathbf{g}(\mathbf{z}_0; \mathbf{p}, \boldsymbol{\theta})$ follow the target posterior $\pi(\Vec{x} \mid \Vec{p})$, we minimize the Kullback-Leibler (KL) divergence between the true posterior $\pi(\Vec{x} \mid \Vec{p})$ and $q_{\theta}$ for any possible observation of pressure $\mathbf{p}$, targeting neural network parameters $\hat{\boldsymbol{\theta}}$ that fulfill the objective
\begin{align}
\widehat{\boldsymbol{\theta}} & =\underset{\boldsymbol{\theta}}{\operatorname{argmin}} \, \mathbb{E}_{\pi(\mathbf{p})}\left[\mathbb{K} \mathbb{L}\left(\pi(\mathbf{x} \mid \mathbf{p}) \| q_{\boldsymbol{\theta}}(\mathbf{x} \mid \mathbf{p})\right)\right] \\
& =\underset{\boldsymbol{\theta}}{\operatorname{argmin}}  \, \mathbb{E}_{\pi(\mathbf{p})}\left[\mathbb{E}_{\pi(\mathbf{x} \mid \mathbf{p})}\left[\log \pi(\mathbf{x} \mid \mathbf{p})-\log q_{\boldsymbol{\theta}}(\mathbf{x} \mid \mathbf{p})\right]\right] \label{eq:KL_div}
\end{align}
Since the $\log$ posterior density $\log \pi( \mathbf{x} \mid \mathbf{p})$ does not depend on $\boldsymbol{\theta}$, \eqref{eq:KL_div} can be further simplified to
\begin{align}
\widehat{\boldsymbol{\theta}} & =\underset{\boldsymbol{\theta}}{\operatorname{argmax}}  \, \mathbb{E}_{\pi(\mathbf{p})}\left[\mathbb{E}_{\pi(\mathbf{x} \mid \mathbf{p})}\left[\log q_{\boldsymbol{\theta}}(\mathbf{x} \mid \mathbf{p})\right]\right] \label{eq:maximum_likelihood}\\
& =\underset{\boldsymbol{\theta}}{\operatorname{argmax}} \, \iint \pi(\mathbf{p}, \mathbf{x}) \log q_{\boldsymbol{\theta}}(\mathbf{x} \mid \mathbf{p}) d \mathbf{p} d \mathbf{x}.
\end{align}
This optimization seeks parameters that maximize the log-likelihood of observing $\mathbf{x}$ given $\mathbf{p}$ for all combinations of $\mathbf{x}$ and $\mathbf{p}$. Since $\mathbf{g}^{-1}(\mathbf{x} ; \mathbf{p}, \boldsymbol{\theta})=\mathbf{z}$ by design, using the change of variables formula leads to
\begin{equation}
q_\theta(\mathbf{x} \mid \mathbf{p})=\pi\left(\mathbf{z}=\mathbf{g}^{-1}(\mathbf{x} ; \mathbf{p}, \boldsymbol{\theta})\right)\left|\left(\frac{d \mathbf{g}^{-1}(\mathbf{x} ; \mathbf{p}, \boldsymbol{\theta})}{d \mathbf{x}}\right)\right| .
\end{equation}
Thus, we can rewrite our objective as
\begin{align}
\widehat{\boldsymbol{\theta}} &= \underset{\boldsymbol{\theta}}{\operatorname{argmax}} \iint \pi(\mathbf{p}, \mathbf{x}) \log q_{\boldsymbol{\theta}}(\mathbf{x} \mid \mathbf{p}) d \mathbf{p} d \mathbf{x} \\
&= \underset{\boldsymbol{\theta}}{\operatorname{argmax}} \iint \pi(\mathbf{p}, \mathbf{x})\left(\log \pi\left(\mathbf{g}^{-1}(\mathbf{x} ; \mathbf{p}, \boldsymbol{\theta})\right) \right. \nonumber \\
&\qquad \qquad \qquad \left. + \log \left|\boldsymbol{J}_{\mathbf{g}^{-1}}\right|\right) d \mathbf{p} d \mathbf{x}, \label{eq:analytical_max_likelihood}
\end{align}
where we have abbreviated $d  \mathbf{g}^{-1}(\mathbf{x} ; \mathbf{p}, \boldsymbol{\theta}) / d \mathbf{x}$ (the Jacobian of $\mathbf{g}^{-1}$ evaluated at $\mathbf{x}$ and $\mathbf{p}$ ) as $\boldsymbol{J}_{\mathbf{g}^{-1}}$. Due to the architecture of the CINN, the $\log \left|\boldsymbol{J}_{\mathbf{g}^{-1}}\right|$ is easy to compute (see Sect. \ref{subsec:coupling_transform}).

Given a joint set of $D$ sound field samples \( \left\{\mathbf{x}_d, \mathbf{p}_d\right\}_{d=1}^{D} \) we can approximate the expectations by minimizing the Monte Carlo estimate of the negative of \eqref{eq:analytical_max_likelihood}, as delineated by
\begin{align}\label{eq:max_likelihood_loss2}
    \widehat{\boldsymbol{\theta}} & =\underset{\boldsymbol{\theta}}{\operatorname{argmin}} \frac{1}{D} \sum_{d=1}^{D} -\log q_\theta\left(\mathbf{x}_d \mid \mathbf{p}_d \right) \\
                                  & =\underset{\boldsymbol{\theta}}{\operatorname{argmin}} \frac{1}{D} \sum_{d=1}^{D} \left( -\log \pi\left(\mathbf{g}^{-1}(\mathbf{x}_{d} ; \mathbf{p}_{d}, \boldsymbol{\theta})\right) - \log \left|\boldsymbol{J}_{\mathbf{g}^{-1}}^{(d)}\right| \right) \\
                                  & = \underset{\boldsymbol{\theta}}{\operatorname{argmin}} \frac{1}{D} \sum_{d=1}^{D} \left(\frac{\left\|\mathbf{g}^{-1}\left(\mathbf{x}_{d} ; \mathbf{p}_{d},\boldsymbol{\theta}\right)\right\|_{2}^{2}}{2} - \log \left|\mathbf{J}_{\mathbf{g}^{-1}}^{(d)}\right| \right). \label{eq:NLL}
\end{align}
By minimizing \eqref{eq:NLL}, the CINN acts as a a conditional density estimator and can efficiently parameterize any sound field through amortized Bayesian inference (see Sec. \ref{subsec:SF_reconstruction}). The graphical representation of the CINN is outlined in Fig. \ref{fig:flows_depiction}, which displays both forward and inverse directions of the CINN.
\newcommand{\captionzero}{Probabilistic graph of a CINN composed of multiple normalizing flows $\mathbf{g}_0 \dots \mathbf{g}_{N_f-1}$ and their inverses $\mathbf{g}_1^{-1} \dots \mathbf{g}_{N_f}^{-1}$ where the variables $\mathbf{z}_{0}, \mathbf{z}_{1}, \dots, \mathbf{z}_{N_f-1}$ are latent variables, $\mathbf{x}$ are the plane wave coefficients, and $\mathbf{p}$ is a conditional variable, used as an auxiliary input of sound pressure to normalizing flows}

\begin{figure}[!t]
\centering
\includegraphics[width=\columnwidth]{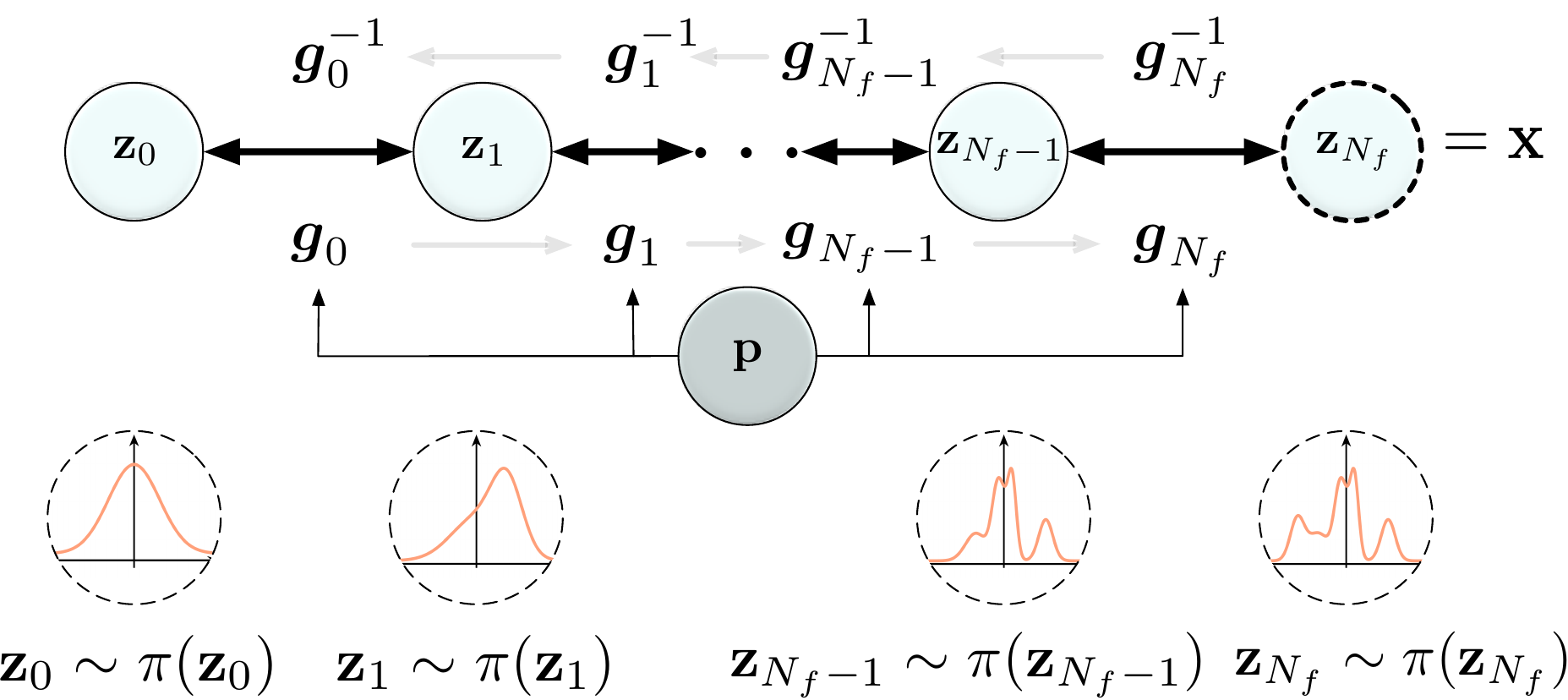}
\caption{\captionzero}
\label{fig:flows_depiction}
\end{figure}

To facilitate interactions among input variables in normalizing flows, each flow is randomly shuffled through permutations, an invertible linear transformation with an absolute determinant equal to 1.\cite{Kingma2018} Specifically, these transformations are implemented through an LU decomposition for each flow
\begin{equation}
    \Vec{z}_{n + 1} =\mathbf{P L} \mathbf{U} \boldsymbol{g}_{n}(\Vec{z}_{n}; \Vec{p}, \boldsymbol{\theta})
\end{equation}
where $\mathbf{L}$ is lower triangular with ones on the diagonal, $\mathbf{U}$ is upper triangular with non-zero diagonal entries, and $\mathbf{P}$ is a permutation matrix. The determinant of the transformation is readily obtained by multiplying the diagonal elements of the upper matrix $\mathbf{U}$ of the decomposition. Moreover, the inverse transformation can be executed by solving two triangular systems, one for $\mathbf{U}$ and one for $\mathbf{L}$. More details are found in \cite{Kingma2018}.

We adopt a strategy where the model is fitted to a Monte-Carlo simulation of random wave fields \cite{FernandezGrande2023, Karakonstantis2023}. Simulating the forward model of wave propagation allows the network to learn the summary statistics of the underlying process \cite{Cranmer2020}. In this particular instance and similar to \eqref{eq:planewaveexpansion}, the sound fields are a linear combination of $N$ plane waves with randomly sampled coefficients, such that the pressure at fixed positions is given by 
\begin{equation}
\Vec{p}_d = \Mat{H} \Vec{x}_d + \Vec{n}_d,
\end{equation}
where $\Vec{x}_i$ and $\Vec{n}_i$ are random variables given by
\begin{equation}\label{eq:training_data_coeffs}
    \pi_x(\Vec{x}_d | \sigma_x) = \mathcal{N}_c (\Vec{x}_i | 0, \sigma_x^2 \Vec{I}),
\end{equation}
and
\begin{equation}\label{eq:training_data_noise}
\pi_n( \mathbf{n}_d | \sigma_p) = \mathcal{N}_c\left(\mathbf{n}_d | 0. \sigma_p^2 \mathbf{I} \right),
\end{equation}
%
The hyperprior \( \sigma_p \) is determined by
\begin{equation}
    \sigma_p^2 = \frac{\mathbb{E}\left[ \check{\Vec{p}}_d^2 \right]}{\operatorname{SNR}},
\end{equation}
where $\check{\Vec{p}}_i = \Mat{H} \Vec{x}_d$ and the variable SNR is uniformly distributed with lower bound $\text{SNR}_l$ and upper bound $\text{SNR}_h$, both in decibels, given by
\begin{equation}
    \pi( \operatorname{SNR}) = \mathcal{U}\left(10^{\left( \text{SNR}_l / 10 \right)}, 10^{\left( \text{SNR}_h / 10 \right)}  \right).
\end{equation}
\subsection{Conditional coupling transform}\label{subsec:coupling_transform}
In practice, CINNs can use a variety of bijective normalizing flows—such as linear, element-wise, planar, radial, and autoregressive transformations—to model complex distributions \cite{Kobyzev2021}. Among these, the coupling transformation stands out for its straightforward yet versatile approach to handling high-dimensional data \cite{Dinh2017}. This method begins by dividing the input vector $\mathbf{z} \in \mathbb{C}^{N}$ into two segments $\mathbf{z}^A \in \mathbb{C}^{\mathrm{n}}$ and $\mathbf{z}^B \in \mathbb{C}^{N-\mathrm{n}}$. The segment $\mathbf{z}^B$ is then merged with a corresponding pressure vector $\Vec{p} \in \mathbb{C}^{M}$, forming a new vector $\Vec{u}^B \in \mathbb{C}^{N + M-\mathrm{n}}$. Through a neural network $\mathrm{NN}$, parameterized by $\boldsymbol{\psi}$, a set of parameters $\boldsymbol{\phi}$ is derived for the invertible function $\boldsymbol{h}: \mathbb{C}^n \rightarrow \mathbb{C}^n$. The transformation, defined as $\mathbf{y} = \boldsymbol{g}\left(\mathbf{z} ; \mathbf{p}, \boldsymbol{\theta}\right)$, is given by
\begin{align} 
    \boldsymbol{\phi}_1, \dots, \boldsymbol{\phi}_n &= \mathrm{NN}\left(\mathbf{u}^B ;\boldsymbol{\psi} \right), \label{eq:coupling_transformA}\\
    \mathbf{y} &= \left[ \left[h_1\left(z_1^A ; \boldsymbol{\phi}_1\right), \ldots, h_{\mathrm{n}}\left(z_{\mathrm{n}}^A ; \boldsymbol{\phi}_{\mathrm{n}}\right) \right]^T, \mathbf{z}^B \right]^T, \label{eq:coupling_transformB}
\end{align}
where each $h_i\left(\cdot; \boldsymbol{\phi}_i\right): \mathbb{C} \rightarrow \mathbb{C}$ represents a scalar bijection. 

The resulting Jacobian matrix of this transform is block triangular,
\begin{align}
\mathbf{J}_{\boldsymbol{g}} &= \left[\begin{array}{ll}
\partial \boldsymbol{y}^A / \partial \boldsymbol{z}^A & \partial \boldsymbol{y}^A / \partial \boldsymbol{z}^B \\
\partial \boldsymbol{y}^B / \partial \boldsymbol{z}^A & \partial \boldsymbol{y}^B / \partial \boldsymbol{z}^B
\end{array}\right] \nonumber \\
&= \left[\begin{array}{cc}
\mathbf{J}_{\boldsymbol{h}} & \partial \boldsymbol{y}^A / \partial \boldsymbol{z}^B \\
\mathbf{0} & \mathbf{I}
\end{array}\right], 
\end{align}
where $\mathbf{J}_{\boldsymbol{h}}$ denotes the Jacobian of the transformed $\mathbf{z}^A$ components, and $\mathbf{I}$ is the identity matrix corresponding to the unchanged $\mathbf{z}^B$ components, while $\boldsymbol{y}^A$ and $\boldsymbol{y}^B$ are the respective partioned outputs of the transform. Achieving exact invertibility in a single pass involves reversing the described process and utilizing $\boldsymbol{g}^{-1}$, $\boldsymbol{h}^{-1}$ and $\mathbf{y}$ in place of $\boldsymbol{g}$, $\boldsymbol{h}$, and $\mathbf{z}$ \cite{Dinh2015}.

%
\subsection{Rational quadratic splines} \label{subsec:Splines}
In principle, the normalizing flows \(\boldsymbol{g}\) have the flexibility to assume any parametric and invertible form. Among these, rational quadratic splines (RQS) are particularly noteworthy. They offer a balance of gradient stability throughout the learning process and versatility in modeling \cite{Durkan2019}. Rational quadratic splines are functions composed of segments (bins) representing the ratio of two quadratic polynomials. The segments are delimited by $K+1$ coordinates known as knots $\left\{\left(z^{(k)}, y^{(k)}\right)\right\}_{k=0}^K$. For instance, in the \(i\)-th dimension of the input $\mathbf{z}^A$ which is binned in the \(k\)-th segment of the spline, the element-wise functions of \eqref{eq:coupling_transformB} take the form
\begin{equation}
    h_i\left(z_i; \phi_i\right) = \frac{\alpha^{(k)}(\xi)}{\beta^{(k)}(\xi)},
\end{equation}
where \(\alpha^{(k)}\) and \(\beta^{(k)}\) are two polynomials, whose parameters are determined by a neural network and $\xi(z_i)=\left(z_i -z_i^k\right) /\left(z_i^{k+1}-z_i^k\right)$ represents the normalized distance along the $z$ axis within the $k^{\text{th}}$ bin. Therefore, $\xi(z_i) \in [ 0, 1]$. The spline is constructed within the specified bounding box \([-B, B]\times[-B, B]\), with the identity function applied elsewhere, while $B$ is typically determined before training and set to a value that covers the range of the input data $\mathbf{x}$. More details on the definition of RQS can be found in Appendix \ref{section:appendixA}.

Following \cite{Durkan2019}, the implementation of the monotonic rational quadratic coupling transform can be outlined by the process below.

\begin{algorithmic}[1]
\STATE A neural network takes the combined vector of pressure and coefficients $\mathbf{u}^B$ as input (see \eqref{eq:coupling_transformA} Sec. \ref{subsec:coupling_transform}) and generates an unconstrained parameter vector $\boldsymbol{\phi}_i$ for each $i=1, \ldots, n$. Each $\boldsymbol{\phi}_i$ has a length of $3K-1$.
\STATE Divide $\boldsymbol{\phi}_i$ into three parts: $\boldsymbol{\phi}_i^w$, $\boldsymbol{\phi}_i^h$, and $\boldsymbol{\phi}_i^d$. $\boldsymbol{\phi}_i^w$ and $\boldsymbol{\phi}_i^h$ have a length of $K$, while $\boldsymbol{\phi}_i^d$ has a length of $K-1$.
\STATE Apply softmax functions to $\boldsymbol{\phi}_i^w$ and $\boldsymbol{\phi}_i^h$, followed by scaling by $2B$. These outputs represent the widths and heights of the $K$ bins, ensuring positivity and coverage of the interval $[-B, B]$.
\STATE The cumulative sums of the widths and heights yield the $K+1$ knots.
\STATE $\boldsymbol{\phi}_i^d$ undergoes a softplus transformation, representing the values of the derivatives at the internal knots.
\end{algorithmic}

Evaluating the rational-quadratic spline transform at a given location \( x \) involves efficiently determining the corresponding bin, typically achieved through binary search due to sorted bins. The Jacobian determinant is computed as a closed-form product of quotient derivatives, while the inversion process entails solving a quadratic equation, the coefficients of which depend on the target value.

Unlike simpler transformations, which offer limited flexibility, a differentiable RQS with sufficient bin count can approximate any monotonic function within the specified interval \( [-B, B]^2 \). Moreover, due to the closed-form, tractable Jacobian determinant it can be analytically inverted.

\subsection{Reconstructing posterior sound fields via amortized Bayesian inference} \label{subsec:SF_reconstruction}
In order for sound field models to be useful in practice, their parameter estimation should be feasible within reasonable time limits. Amortized Bayesian inference aims to efficiently map the relationship between observations and posterior distributions through a conditional density estimator such as the trained CINN \cite{Cranmer2020}. The initial cost of training the CINN is `amortized', as inference on new observations of pressure $\Vec{p}$ becomes more efficient. This efficiency is due to the ability to achieve posterior estimates $\mathbf{x} \sim q_{\boldsymbol{\theta}}( \Vec{x} \mid \Vec{p})$ via a single forward pass through the CINN.

Therefore, sampling from an approximate posterior involves first drawing samples from the latent distribution \( \mathbf{z}_0 \sim \mathcal{N}(\mathbf{z}_0 \mid \mathbf{0}, \mathbf{I}) \) and then passing them through the function \( \Vec{g} \left(\mathbf{z}; \mathbf{p},\widehat{\boldsymbol{\theta}}\right)  \), as depicted in Fig. \ref{fig:flows_depiction}. This sampling mechanism is highly efficient, typically requiring only a fraction of a second to complete.

Since the prior over coefficients \( \mathbf{x} \) follows a normal distribution, e.g., \eqref{eq:training_data_coeffs}, one can estimate the exact MAP coefficients \( \widehat{\mathbf{x}} \) by first obtaining the latent variable \( \widehat{\mathbf{z}} \) which maximizes the posterior probability \( \pi(\mathbf{x} | \mathbf{p}) \). The likelihood is given by
\begin{align}
    \pi(\mathbf{p}|\mathbf{x}) &= \pi_z \left(\mathbf{p}|\mathbf{z}; \widehat{\boldsymbol{\theta}} \right) \nonumber \\
    &= \frac{1}{\sqrt{2 \pi \sigma_{p}^2}} \exp \left[-\frac{\left(\mathbf{p}-\mathbf{H} \cdot \Vec{g}\left( \mathbf{z};\mathbf{p}, \widehat{\boldsymbol{\theta}} \right)\right)^2}{2 \sigma_{p}^2}\right],
\end{align}
where \( \sigma_{p}^2 \) represents the variance in the measurement noise. Therefore, The MAP coefficients are determined as the solution to the optimization problem
\begin{align}\label{eq:CINN_MAP}
    \widehat{\Vec{x}}
    &= \Vec{g} \left(\widehat{\mathbf{z}};\Vec{p}, \widehat{\boldsymbol{\theta}} \right) = \underset{\Vec{x}}{\operatorname{argmax}} \left( \operatorname{log}\pi \left(\Vec{x} | \Vec{p} \right) \right) \nonumber \\
    &= \underset{\Vec{z}}{\operatorname{argmin}} \Bigg[\frac{\normtwo{\Vec{p}-\Vec{H} \cdot \Vec{g}\left( \Vec{z}; \Vec{p}, \widehat{\boldsymbol{\theta}} \right)}}{2 \sigma_{p}^2} \nonumber \\ 
    &+ \frac{1}{2 \sigma_z^2} \normtwo{\Vec{z}} + \frac{1}{2 \sigma_x^2} \normtwo{\Vec{g}\left( \Vec{z}; \Vec{p}, \widehat{\boldsymbol{\theta}} \right)}\Bigg],
\end{align}
with $\sigma_z = 1$. 

Algorithm \ref{alg:CINN} outlines the training and inference process of the CINN. The procedure is divided into two main phases: training and inference. During the training phase, the algorithm iteratively samples wave coefficients and noise from their respective prior distributions, simulates pressure fields, and updates the neural network parameters by minimizing a loss function until convergence is achieved. The second phase describes the inference stage which, depending on the specific goal, can either compute the MAP estimates as described in \eqref{eq:CINN_MAP} or generate samples from the approximate posterior distribution \eqref{eq:CINN_approx_posterior}, providing insights into the underlying uncertainty.
\begin{algorithm}
\caption{CINN for amortized Bayesian inference}
\label{alg:CINN}
\begin{algorithmic}[1]
\STATE \textbf{Training Phase} (\textit{Batch Size} $D$):
\REPEAT
    \FOR{$d=1$ \TO $D$}
        \STATE Sample coefficients from prior: $\mathbf{x}_d \sim \pi_x(\mathbf{x}_d | \sigma_x)$.
        \STATE Sample signal-to-noise ratio (SNR) from its prior: $\text{SNR} \sim \pi(\text{SNR})$.
        \STATE Define noise variance based on SNR: $\sigma_p^2 = \frac{\mathbb{E}\left[\|\Vec{p}_d\|^2\right]}{\text{SNR}}$.
        \STATE Sample noise from its distribution: $\Vec{n}_d \sim \pi_n(\Vec{n}_d | \sigma_p)$.
        \STATE Simulate the pressure field: $\Vec{p}_d = \Mat{H}\Vec{x}_d + \Vec{n}_d$.
        \STATE Pass the data $\{\Vec{x}_d, \Vec{p}_d\}$ through the inverse CINN: $ \Vec{z}_d = \mathbf{g}^{-1}(\Vec{x}_d; \Vec{p}_d, \boldsymbol{\theta})$.
        \STATE Compute the loss according to \eqref{eq:NLL} and obtain gradients w.r.t. $\boldsymbol{\theta}$.
        \STATE Update $\boldsymbol{\theta}$ using backpropagation.
    \ENDFOR
\UNTIL{convergence to $\widehat{\boldsymbol{\theta}}$}
\newline
\STATE \textbf{Inference Phase} (\textit{Experimental pressure} $\Vec{p}$):
\IF{estimate $\underset{\Vec{x}}{\text{argmax}}\; \pi(\Vec{x} | \Vec{p})$}
    \STATE Sample a latent variable: $\Vec{z} \sim \mathcal{N}(\Vec{z} \mid \Vec{0}, \mathbf{I})$.
    \REPEAT
        \STATE Pass $\Vec{p}$ and $\Vec{z}$ through the forward CINN: \\ $\Vec{x} = \mathbf{g}(\Vec{z}; \Vec{p}, \widehat{\boldsymbol{\theta}})$.
        \STATE Compute loss according to \eqref{eq:CINN_MAP} and obtain gradients w.r.t. $\Vec{z}$.
        \STATE Update $\Vec{z}$ using backpropagation
    \UNTIL{convergence to $\widehat{\Vec{x}}$}
    \STATE Return $\widehat{\Vec{x}} = \underset{\Vec{x}}{\text{argmax}}\; \pi(\Vec{x} | \Vec{p})$.
    
\ELSIF{sample from $q_{\boldsymbol{\theta}}(\Vec{x} | \Vec{p})$}
    \FOR{$l=1$ \TO $L$ samples}
        \STATE Sample a latent variable: $\Vec{z}_l \sim \mathcal{N}(\Vec{0}, \mathbf{I})$.
        \STATE Pass $\Vec{p}$ and $\Vec{z}_l$ through the forward CINN: \\ $\Vec{x}_l = \mathbf{g}(\Vec{z}_l; \Vec{p}, \widehat{\boldsymbol{\theta}})$.
    \ENDFOR
    \STATE Return $\{\Vec{x}_l\}_{l=1}^L$ as samples from the approximate posterior $q_{\boldsymbol{\theta}}(\Vec{x} | \Vec{p})$.
\ENDIF
\end{algorithmic}
\end{algorithm}
\subsection{Hierarchical Bayes for inference and inversion}\label{subsec:hierarch_bayes}
We design a hierarchical plane wave model with sparsity-inducing priors to compare with the proposed CINN architecture. The hierarchical nature of the model allows for the incorporation of multiple layers of abstraction, each layer refining and focusing on the information passed from the previous one, thus enabling a more nuanced understanding of the sound field. The assignment of sparsity-inducing priors is not merely a computational convenience but a reflection of the physical reality that many sound fields are inherently sparse in their representation, particularly at low frequencies \cite{Mignot2013,Verburg2018}. 

Assuming the noise ${\bf n}\sim\mathcal{N}_{c}\left({\bf n} | \mathbf{0}, \sigma_{p}^2 \mathbf{I} \right)$ with $\sigma_{p}^{2}$  expressing variance of model, then the likelihood is a product of complex Gaussians
\begin{equation}
\pi(\mathbf{p}|\mathbf{x};\sigma_{p}) =\prod_{m=1}^{M} \mathcal{N}_{c}\left(p_m | \mathbf{H}_m \mathbf{x}, \sigma_{p}^2 \mathbf{I} \right).
\end{equation}
%
Prior probabilities over the coefficients can be assigned hierarchically, emphasizing the selection of weakly informative priors. In particular, a complex normal distribution is selected to represent each heteroscedastic complex coefficient such that 
\begin{equation}
    \pi(\Vec{x} | \boldsymbol{\sigma}_{x}) = \prod_{n=1}^{N}\mathcal{N}_c(x_n |0, \sigma_{x_n}^2).
\end{equation}
The hyper-priors $\sigma_{x}^2$ are selected to be weakly informative, as they put more probability mass on lower values, and are selected such that there is an individual variance associated independently with every coefficient
\begin{equation}
    \pi(\boldsymbol{\sigma}_{x}^2 ; \alpha, \beta) = \prod_{n=1}^{N}\mathcal{IG}(\sigma_{x_n}|\alpha, \beta),
\end{equation}
where $\mathcal{IG}(\cdot)$ is an inverse Gamma distribution, with $\alpha$ and $\beta$ referring to the shape and scale parameters respectively. This selection of variances also aligns with the sparse Bayesian learning principles, such that the selection of priors is independent of the scale of the measured data (e.g., scale-invariance) \cite{Tipping2001}. 
The noise in the model is expressed via the hyper-prior $\sigma_{p}$, which in turn is estimated by the standard deviation of the noise floor, obtained during the experimental process. Therefore, the posterior distribution is given by
%
\begin{equation}\label{eq:hierarchic_posterior}
    \pi(\mathbf{x}|\mathbf{p}, \boldsymbol{\sigma}_{x}, \sigma_{p}) = \frac{\pi(\mathbf{p}|\mathbf{x};\sigma_{p}) \pi(\mathbf{x}|\boldsymbol{\sigma}_{x})\pi(\boldsymbol{\sigma}_{x}^2;\alpha, \beta)}{\pi(\mathbf{p})}.
\end{equation}
The complete hierarchical model is depicted in Fig. \ref{fig:plane_wave_pgm}. The posterior probability \eqref{eq:hierarchic_posterior} is often approximated by using MCMC sampling methods, \cite{Rubin1995} a class of algorithms for numerically approximating high-dimensional integrals.
\newcommand{\captionzeroone}{Probabilistic graph of hierarchical plane wave model}

\begin{figure}[!t]
\centering
\includegraphics[width=.7\columnwidth]{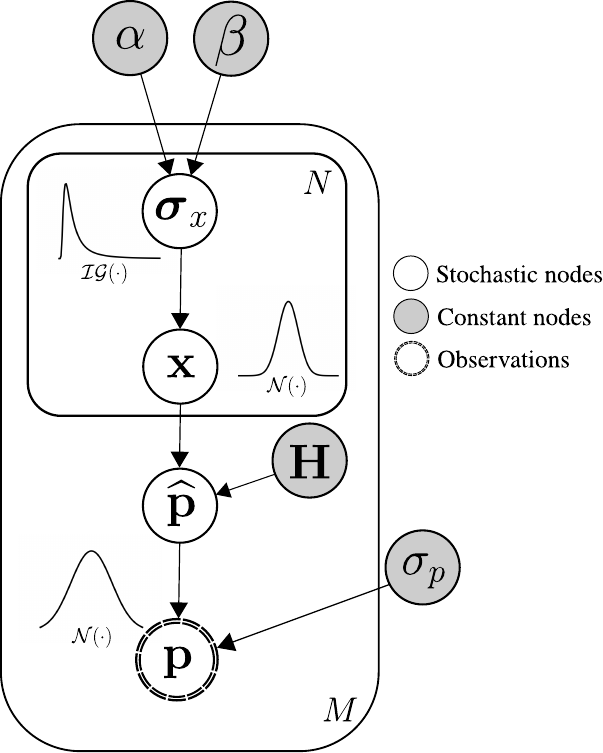}
\caption{\captionzeroone}
\label{fig:plane_wave_pgm}
\end{figure}

The maximum a posteriori (MAP) estimate of the coefficients \(\widehat{\Vec{x}}\) is determined by
\begin{equation}\label{eq:MAP}
\widehat{\Vec{x}} \propto \operatorname{argmax}{\pi(\mathbf{p}|\mathbf{x}; \sigma_{p})\pi(\mathbf{x}|\boldsymbol{\sigma}_{x})\pi(\boldsymbol{\sigma}_{x}^2;\alpha, \beta)},
\end{equation}
where the evidence $\pi(\Vec{p})$ remains constant across the unknown coefficients.

Challenges arise when one requires the complete distribution as described by \eqref{eq:hierarchic_posterior}, due to the computational demands of evaluating the likelihood function, heightened by the complexity of the underlying physical system which requires numerous evaluations of the forward model. Moreover, the high-dimensional nature of the parameters $\mathbf{x}$ complicates the derivation of posterior samples, underscoring the need for efficient computational strategies in Bayesian inference. These are common issues with sampling from high-dimensional posteriors using MCMC algorithms. 


\section{Experimental dataset}\label{sec:experimental_dataset}
Acoustic measurements were conducted in Auditorium A of the Niels Bohr Institute in Copenhagen, Denmark, a renowned Danish scientific facility. Largely preserved in its original state, Auditorium A has been a venue for numerous historical lectures and pioneering talks on quantum mechanics. Figure \ref{fig:NBI_data} illustrates the auditorium layout, along with images depicting the measurement process.

The auditorium is equipped with large windows on the wall on the southeast side (at $x = 0$ m), complemented by light drapery, while concrete walls enclose the remaining sections. Wooden benches, arranged for seating during lectures, occupy a significant portion of the auditorium's length (spanning $y \in [0, 6.3]$ m). The floor of the auditorium is slanted, progressively reducing the room height at each successive bench section toward the rear. Table \ref{tab:room_char} provides details on the auditorium's reverberation time and volume. Room impulse responses (RIRs) were measured using a Universal Robots UR5 robotic arm equipped with a 1/2" omnidirectional Br{\"u}el \& Kjær pressure microphone. The measurements were conducted across a uniform cuboid grid comprising 4913 positions (17 per dimension), with each dimension measuring $|x| = 0.5$ m, $|y| = 0.5$ m, and $|z| = 0.44$ m. RIRs were obtained using the exponential sine sweep and inverse filter method, employing 5-second sweeps. A Dynaudio BM6 2-way passive loudspeaker served as the sound source, positioned at the lecturer's location (front) of the auditorium, while the measurement aperture was situated at the rear, above the last seating row.

For fitting the invertible neural network, only a subset of this data was used, by uniformly subsampling the complete grid to 4 positions per dimension, and also limiting the aperture size to finally obtain an array of dimensions $ \left|x\right| = 0.28125$ m, $\left|y\right| = 0.28125$ m, $\left|z\right| = 0.2475$ m composed of 64 microphones. The rest of the measured RIRs were used for experimental validation. 
\begin{table}[!t]
\renewcommand{\arraystretch}{1.3}
\caption{Size and reverberation times in octave bands from the studied room (Niels Bohr Institute - Auditorium A)}
\label{tab:room_char}
\centering
\resizebox{\columnwidth}{!}{%
\begin{tabular}{|c|cccccc|}
\hline
Volume (m$^3$) & \multicolumn{6}{c|}{Reverberation Time - $T_{60}$ (s)} \\ \hline
\multirow{2}{*}{$134$} & \multicolumn{1}{c|}{125 Hz} & \multicolumn{1}{c|}{250 Hz} & \multicolumn{1}{c|}{500 Hz} & \multicolumn{1}{c|}{1 kHz} & \multicolumn{1}{c|}{2 kHz} & 4 kHz \\ 
\cline{2-7}  & \multicolumn{1}{c|}{0.572}    & \multicolumn{1}{c|}{0.728}   & \multicolumn{1}{c|}{0.797}   & \multicolumn{1}{c|}{0.874}  & \multicolumn{1}{c|}{1.03}  &  1.06  \\ \hline
\end{tabular}
}
\end{table}
\newcommand{\captionone}{Experimental dataset layout in the Niels Bohr Institue - Auditorium A (top) and photographs of the auditorium and the robotic arm used for measuring in the microphone array configuration (bottom)}

\begin{figure}[!t]
\centering
\includegraphics[width=\columnwidth]{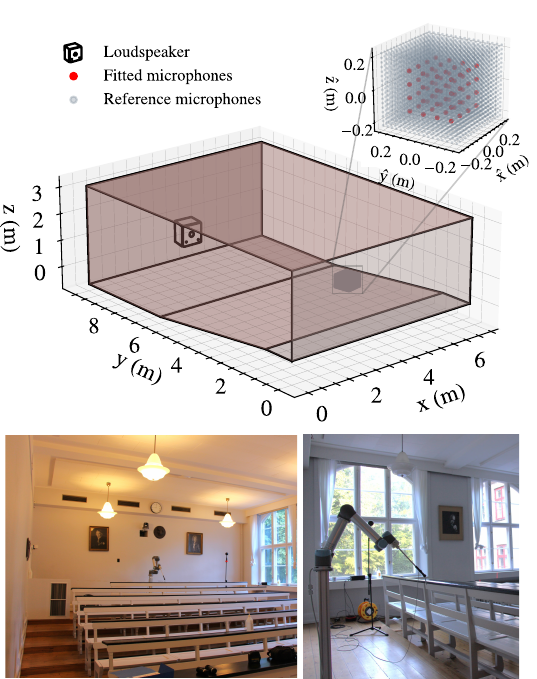}
\caption{\captionone}
\label{fig:NBI_data}
\end{figure}
\section{Results}\label{sec:results}
\subsection{CINN training} \label{subsec:CINN_training}
The training process of the CINN involves conducting Bayesian hyperparameter selection using the Optuna package \cite{Akiba2019} to ensure optimal performance. In our experiments, we set $\sigma_x=0.3$ to determine the prior over the coefficients of the sound fields, while specifying signal-to-noise ratio (SNR) levels ranging from $\text{SNR}_l = 25$ to $\text{SNR}_h = 45$ for the random wave field data. An initial learning rate of $\eta = 10^{-5}$ was employed, coupled with an exponential decay learning rate scheduler using the Adam optimizer \cite{Kingma2017}. The training dataset (and the later experimental validation) consisted of $N = 642$ plane waves, with a batch size of 256 sound fields. The coefficient vectors were split in half when processed by each normalizing flow layer (i.e. $\mathrm{n} = 321$). Finally, to accommodate complex-valued data, we adopt a strategy that involves independently treating the real and imaginary parts of a complex variable. The real and imaginary components of any arbitrary complex variable $\mathbf{v} \in \mathbb{C}^{N_v}$ are concatenated along the same axis, effectively doubling the dimensionality of the variable such that $\mathbf{u} \in \mathbb{R}^{2N_v}$

The CINN was trained jointly on all frequencies of interest, which allows for generalization across different room configurations and frequencies. Each RQS comprised $K=8$ bins, and the CINN was composed of $N_f=12$ flows in total, with random permutations incorporated between flows (except for the last flow). The feed-forward neural networks parameterizing the spline flows were each composed of two layers with 1024 neurons per layer and employed Exponential Linear Unit (ELU) activation functions with a hyperparameter $\alpha_{\text{ELU}}$ value of 1. The implementation is carried out using the publicly available Framework for Easily Invertible Architectures Python package \cite{Ardizzone20182022}.

Training the CINN involved $100000$ iterations on a single NVIDIA A100 80GB GPU, utilizing only 18\% of the system memory. The training duration spanned a total of 16 hours, ensuring thorough convergence. Training validation was conducted by evaluating the averaged normalized mean square error ($\text{NMSE}_{\text{oct}}$) across $N_b = 6$ octave bands between 125 Hz and 4 kHz, of the 64-channel experimental dataset (see Sec. \ref{sec:experimental_dataset}). This evaluation entailed sampling from the CINN at periodic intervals and subsequently averaging the sampled coefficients, yielding the posterior mean for each frequency, and then reconstructing the pressure at the microphone positions. The NMSE and $\text{NMSE}_{\text{oct}}$ are given by 
\begin{align}
    \text{NMSE} &= \frac{||\mathbf{p}-\mathbf{p}_\diamond ||^2}{||\mathbf{p}||^2}, \label{eq:NMSE} \\
    \text{NMSE}_{\text{oct}} &= \frac{1}{N_{b}}\sum_{n=1}^{N_{b}} \text{NMSE}_n , \label{eq:NMSE_with_avg}
\end{align}
where $\mathbf{p}$ refers to the experimental (measured) pressure and $\mathbf{p}_\diamond$ to the estimated pressure, respectively. 
Figure \ref{fig:NLL_NMSE_train} presents both the $\text{NMSE}{\text{oct}}$ (in logarithmic scale) and the negative log-likelihood (NLL), or the logarithm of \eqref{eq:max_likelihood_loss2}, observed during the training process. The $\text{NMSE}_{\text{oct}}$ demonstrates the CINN's capability to fit out-of-distribution data, in this case, that of the experimental observations. Meanwhile, the NLL indicates that the CINN quickly learns to model the forward propagation of plane waves when conditioned on the pressure, even in the early stages of training using the simulated data.
\newcommand{\captionzerotwo}{Validation using $10 \log_{10} \left(\text{NMSE}_{\text{oct}} \right)$ during CINN training (top) and negative log-likelihood (logarithm of \eqref{eq:max_likelihood_loss2}) as training loss on simulated random wave fields (bottom)}

\begin{figure}[!t]
\centering
\includegraphics[width=\columnwidth]{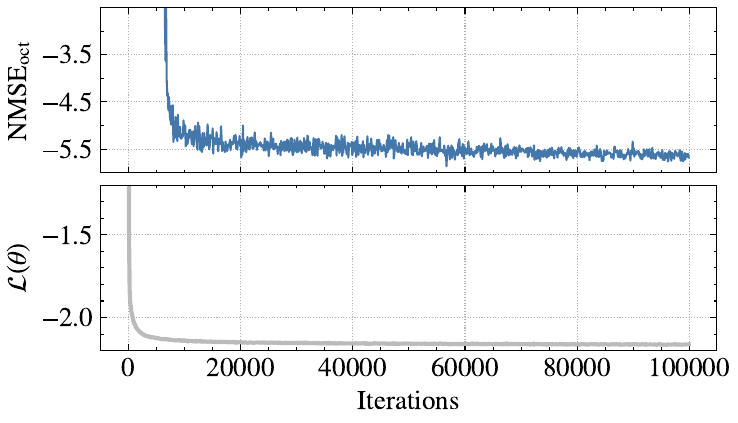}
\caption{\captionzerotwo}
\label{fig:NLL_NMSE_train}
\end{figure}
\subsection{Single plane wave posterior estimation with CINN}
To assess the performance of the proposed invertible neural network, we conduct an investigation focused on estimating the posterior distribution of single-plane wave sound fields. A CINN model is trained using a dataset consisting of sound fields generated by random plane waves emanating from 600 uniformly sampled directions within a 2D circular domain. These sound fields are captured by an $8\times8$ microphone array spanning a $5\times5$ m$^2$ aperture, with amplitudes and phases drawn from distributions outlined in Sec. \ref{subsec:CINN_wave_model}. Training proceeds with the same hyperparameters detailed in Sec. \ref{subsec:CINN_training}, except for the learning rate and number of iterations, adjusted to $\eta = 2 \cdot 10^{-4}$ and 2500, respectively. For MAP inference, as discussed in \eqref{eq:CINN_MAP}, we employ 1000 iterations with a learning rate of $\eta_z = 0.05$ while utilizing the Adam optimizer. The rest of the parameters are determined by the training data.

Following training, we subject the model to an out-of-distribution scenario, exposing it to a simulated single-plane wave sound field characterized by unseen amplitude, phase, and direction, yet sampled using the same microphone array configuration. Figure \ref{fig:toy_model} illustrates the true simulated sound field alongside the reconstructed sound field (MAP estimate) and the element-wise standard deviation of the invertible network.

The model yields promising outcomes, particularly in areas adjacent to the aperture, where the predicted sound field closely matches the ground truth observations. However, discrepancies between the predicted and true sound fields become more evident with increasing distance from the measurement aperture. As expected, this divergence coincides with heightened uncertainty estimates, as indicated by elevated element-wise standard deviations in regions where the model's predictions deviate from observed data. The CINN demonstrates adeptness in quantifying uncertainty, offering valuable insights into areas of prediction discrepancy. While apparent in this straightforward scenario, such findings suggest that CINNs can contribute significantly to enhancing our comprehension of both model and data uncertainties in sound field analysis and signal processing applications, such as sensor placement.
\newcommand{\captiontwo}{Single plane wave reference sound field, MAP inference, and point-wise standard deviation using a CINN; the microphone positions are superimposed over the reference sound field}

\begin{figure}[!t]
\centering
\includegraphics[width=\columnwidth]{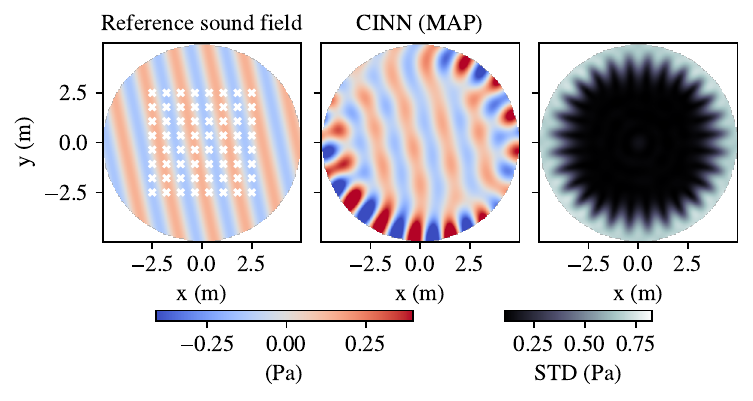}
\caption{\captiontwo}
\label{fig:toy_model}
\end{figure}
\subsection{Quantifying uncertainty in CINN sound field reconstruction}
Quantifying uncertainty in sound field reconstruction using wave-based models is crucial for ensuring the reliability of the results obtained. Given the absence of ground truth sound field data in most practical scenarios, relying solely on a single point estimate may obscure the validity of the solution, underscoring the necessity of incorporating the inherent uncertainty inherent in Bayesian models. This subsection emphasizes the importance of uncertainty quantification in the context of CINN sound field reconstruction when using experimental data, supported by qualitative insights.

Figure \ref{fig:CINN_results} illustrates extrapolated sound fields (volume of 0.11 m$^3$) beyond the array bounds (volume of 0.02 m$^3$) at frequencies of 400 Hz, 800 Hz, 1200 Hz, and 1600 Hz. It juxtaposes the ground truth with the CINN MAP prediction and the point-wise standard deviation, as defined by \eqref{eq:CINN_approx_posterior}, employing a 64-channel microphone array detailed in Sec. \ref{sec:experimental_dataset}. The CINN exhibits the ability to accurately reconstruct the sound fields, even when extrapolating to dimensions nearly twice the size of the measurement array aperture while capturing underlying spatial characteristics across frequencies. Notably, the CINN demonstrates higher confidence in its extrapolated predictions at lower frequencies, with standard deviation values increasing as frequency increases. This observation aligns with the fact that lower frequencies with longer wavelengths exhibit greater spatial coherence, allowing for extrapolation over large distances relative to their wavelength. Conversely, higher frequencies have reduced spatial coherence, leading to increased variability in extrapolation due to the smaller scale of spatial features relative to the wavelength. Additionally, the discernible periodic-like pattern in the standard deviation also suggests this. As the wavelength approaches or becomes smaller than the inter-microphone spacing, the array's ability to unambiguously capture and reconstruct the wavefront diminishes, evident in the array's sensitivity to certain directions. 

\newcommand{\captionfour}{(Color online) Measured and CINN (MAP) extrapolated sound fields with point-wise standard deviation for frequencies $f = 400, \; 800, \; 1200$ and $1600$ Hz; the 64-channel validation microphone array is super-imposed on the first (ground truth) sound field at $f = 300$ Hz} 

\begin{figure*}[ht!]
\centering
\includegraphics[trim=0.1cm 0.12cm 0.1cm 0.11cm, clip,width=\linewidth]{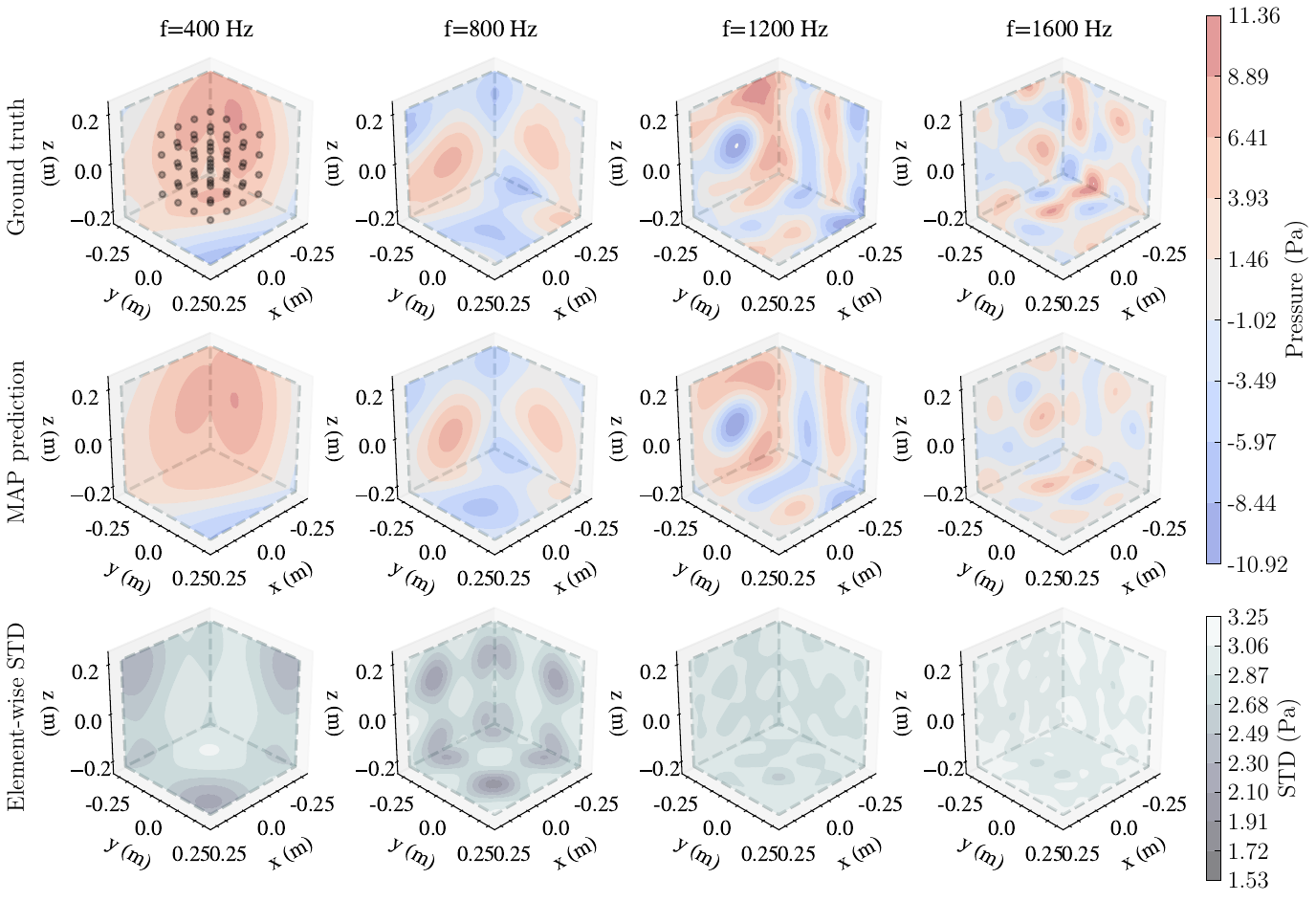}
\caption{\captionfour}
\label{fig:CINN_results}
\end{figure*}
%
\subsection{Comparison of CINN with hierarchical Bayes MCMC for experimental sound field estimation}
The hierarchical Bayes plane wave model parameters are obtained using the gradient-based No-U-Turn sampler (NUTS), which belongs to the class of Hamiltonian Monte-Carlo (HMC) algorithms, chosen for its efficiency, particularly in handling high-dimensional data \cite{Rubin1995}. The Python package Pyro is utilized for MCMC sampling \cite{Bingham2019}. The plane wave dictionary $\Mat{H}$ is also composed of 642 plane waves and the hyperprior $\sigma_p$ is also obtained from the noise floor in the measurements. The rest of the hyperpriors are set to $\alpha = 3$ and $\beta = 1$ accordingly. Model selection was based on the Pareto-smoothed importance sampling leave-one-out cross-validation criteria \cite{Vehtari2022} at $f = 500$ Hz, with the same model used for all other experimental cases. Two chains are generated for each model, each consisting of 1000 warm-up samples and a further 2000 posterior samples, which are then mixed to remove potential non-representative samples. The hierarchical Bayes model is hereby denoted as MCMC for ease of comparison. Each hierarchical model posterior (per frequency) requires approximately 7 minutes per chain. 

On the other hand, sampling from the CINN posterior is very efficient, typically taking only a few seconds to generate thousands of samples. This process involves sampling from the latent distribution followed by a straightforward forward pass through the network. Additionally, recovering the MAP plane wave coefficients, as detailed in \eqref{eq:CINN_MAP}, involves iterating 1500 times with a learning rate of $\eta_z = 0.05$, a task that takes less than a minute on a GPU. Hyperparameter selection, including the aforementioned parameters, is conducted using the Optuna package for Bayesian hyperparameter selection, yielding fairly consistent results across minor variations in parameter settings. 

To evaluate the performance of either method we use the NMSE in decibels as denoted in \eqref{eq:NMSE} and the Spatial Coherence (SC) between the measured pressure $\Vec{p}$ and the predicted pressure $\Vec{p}_\diamond$ given by 
\begin{equation}
     \textsc{SC} = \frac{ \left|\Vec{p}^{H} \Vec{p}_\diamond\right|^2}{(\Vec{p}^{H} \Vec{p}) (\Vec{p}_\diamond^{H} \Vec{p}_\diamond)}, 
 \label{eq:ss}
\end{equation}
where $H$ denotes the Hermitian transpose operation, the SC metric shows the degree of similarity between the pressure fields with values between 0 and 1. 

Figure \ref{fig:NMSE_SpatialCoh} presents the NMSE and SC across frequencies for both the MCMC and CINN models and the 64 microphone array as shown in Fig. \ref{fig:NBI_data}. Specifically for the CINN, results are showcased for both the MAP prediction and the posterior mean prediction (i.e., $\mathbb{E}\left[ q_{\boldsymbol{\theta}}(\mathbf{x} \mid \mathbf{p})\right]$). Additionally, the corresponding metric-wise standard deviations for each method are denoted as $\sigma_{\text{CINN}}$ and $\sigma_{\text{MCMC}}$. The CINN MAP prediction outperforms the respective CINN posterior mean prediction, which is expected given that the CINN models a multivariate distribution across multiple frequencies and is likely not unimodal. Nonetheless, the CINN yields favorable results considering its inference time is negligible, and its adaptability to various types of sound fields. 

Both the CINN MAP estimate and the MCMC MAP estimate exhibit similar performance, with noticeable divergence occurring primarily beyond 1 kHz. This divergence is anticipated, as the hierarchical model can often favor smaller coefficient variances (i.e., $ \norm{\mathbf{x}} \approx 0$) due to the inverse gamma prior assigned over them. Consequently, a degree of sparsity is observed at lower frequencies, but this tendency diminishes at higher frequencies. This behavior is also reflected in the uncertainty in the SC metric for the MCMC method, which is more pronounced at lower frequencies compared to higher frequencies. Employing a less informative prior may enhance performance at higher frequencies, albeit potentially at the expense of performance at lower frequencies. 

The CINN circumvents this model selection issue by relying on data-driven priors. Notably, both CINN MAP and posterior mean point estimates show to be more robust at higher frequencies, reflected in the improvements of both NMSE and SC metrics. Furthermore, the uncertainty associated with the CINN estimates exhibits a decreasing trend in standard deviation with increasing frequency. This trend can be attributed to the CINN's training methodology, which involved random wave models without any sparsity constraints. 
\newcommand{\captionthree}{Performance of CINN and MCMC MAP estimates, as well as the CINN posterior mean and the respective 95 \% confidence intervals of the NMSE (top) and the spatial coherence (bottom), using the experimental 64 channel microphone array}

\begin{figure}[!t]
\centering
\includegraphics[width=\columnwidth]{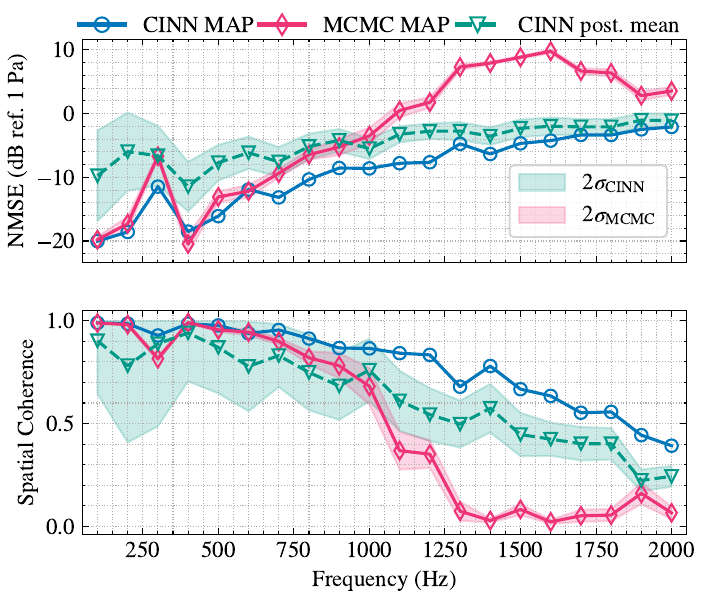}
\caption{\captionthree}
\label{fig:NMSE_SpatialCoh}
\end{figure}

We investigate the robustness of each model concerning the decimation of the microphone array, where the arrays undergo uniform decimation in each dimension, reducing the total number of microphones from 216 to 27. It is important to highlight that the aperture size is adjusted proportionally to ensure a Nyquist rate well above the frequencies under examination.

Figure \ref{fig:Mics_vs_freq} depicts the performance of the posterior means for both the CINN and MCMC methods, with their respective 95\% confidence interval (approximately $2\sigma$). In the case of MCMC, this corresponds to the MAP estimate. The comparison reveals the comparable performance of both methods at low frequencies in terms of NMSE, with the MCMC slightly outperforming the proposed CINN; however, the opposite occurs at high frequencies. 

The CINN demonstrates a slight decrease in NMSE with an increasing number of microphones. Conversely, the MCMC method does not exhibit this behavior, as a larger dataset assigns more weight to the likelihood and less to the prior, particularly when utilizing sparsity-imposing hyperpriors. Consequently, additional data may necessitate recalibration of the hierarchical model.

At low frequencies, the CINN saturates for a large number of microphones, evident at \(f = 250\) Hz for 96 and 216 microphones. Notably, the standard deviation, indicated by the error bars, is significantly larger than the rest of the predictions. This disparity likely stems from variance in the measurement noise present in the experimental data, which may differ from the training data used for the CINN, and is more pronounced at low frequencies.
\newcommand{\captionfive}{Perfomance of CINN and MCMC methods over frequency, each subfigure displays the posterior mean NMSE and 95\% confidence interval at different microphone array configurations}

\begin{figure}[!t]
\centering
\includegraphics[width=\columnwidth]{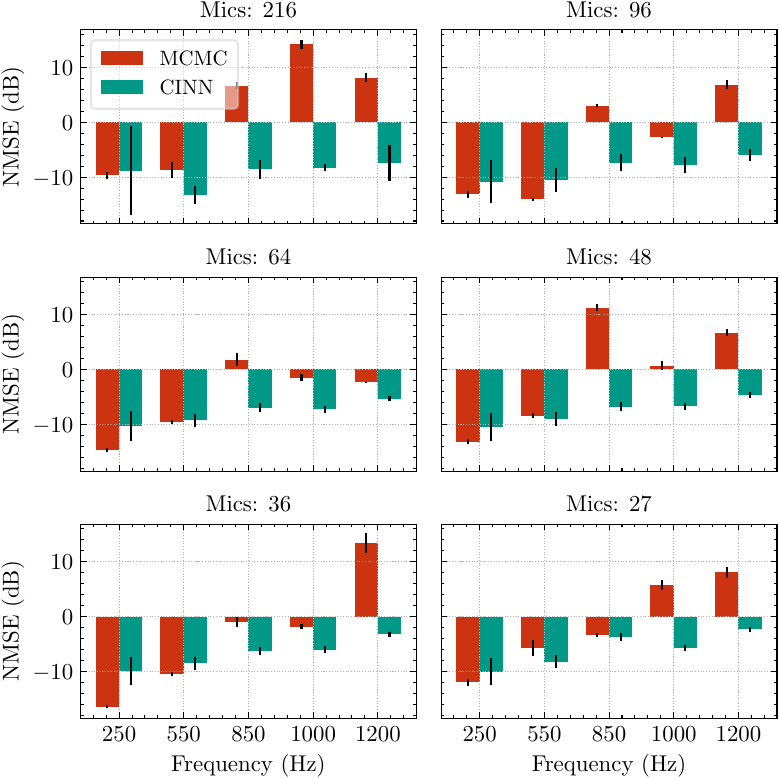}
\caption{\captionfive}
\label{fig:Mics_vs_freq}
\end{figure}

\subsection{Room impulse response reconstruction}
Figure \ref{fig:InterpRIR} demonstrates an example of an interpolated RIR (i.e., within the array bounds), with the top subfigure depicting the CINN MAP prediction and the bottom showing the posterior mean prediction within its respective 99.7\% confidence interval, both superimposed over the measured RIR. It is noteworthy that while the MAP requires an optimization step as outlined in \eqref{eq:CINN_MAP}, its balance of accuracy and efficiency renders it an appealing method for RIR reconstruction. On the other hand, even if not as accurate, obtaining the posterior mean RIR is nearly instantaneous. As mentioned, it merely necessitates sampling from the latent distribution and performing as many forward passes of the network as there are discrete frequencies. This is significant as RIR reconstruction entails a heavy computational load and is a time-consuming task, an aspect often overlooked in the literature.
\newcommand{\captionsix}{Normalized CINN MAP estimated RIR (top) and posterior mean RIR and 99.7 \% confidence interval (bottom) at an interpolated position both superimposed over the measured reference RIR}
\begin{figure*}[ht!]
\centering
\includegraphics[trim=0.1cm 0.12cm 0.1cm 0.11cm, clip,width=\linewidth]{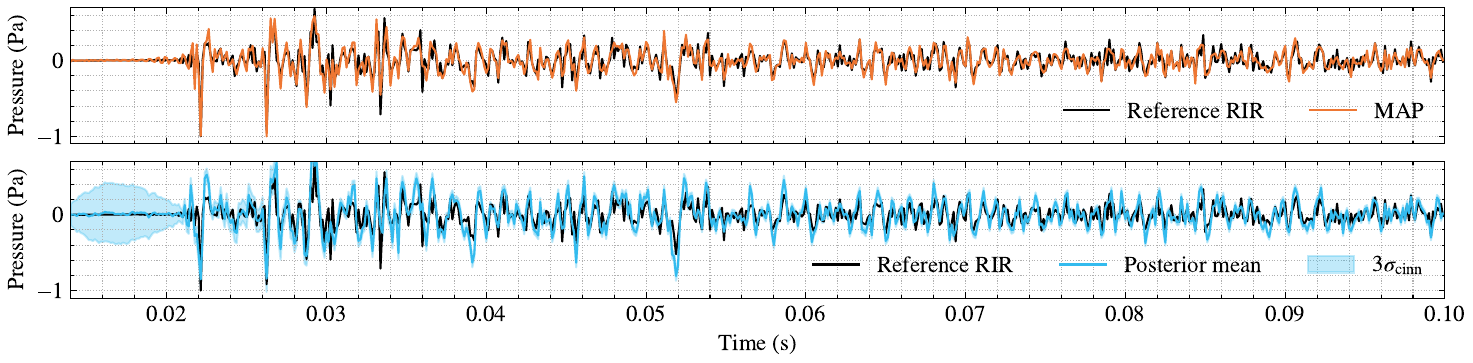}
\caption{\captionsix}
\label{fig:InterpRIR}
\end{figure*}
Figure $\ref{fig:ExtrapRIR}$ displays the reconstructed MAP and posterior mean RIRs at a distance of 0.12 m from the aperture. The MAP estimate demonstrates noteworthy predictive capability, adeptly representing the numerous dense reflections inherent in this dataset, with most peaks aligning well in both magnitude and phase. Conversely, the posterior mean exhibits less precision in terms of magnitude, as the peaks extend beyond the reference RIR when normalized. However, it can correctly predict the alignment of individual reflections with respect to the reference RIR.

The majority of uncertainty in either reconstructed RIR is concentrated at the beginning of the responses, before the direct part, but also encompasses the early reflections. This phenomenon can be attributed to pre-ringing, a characteristic of frequency domain methods, which aligns with the hypothesis of plane wave underrepresentation, potentially inducing this phenomenon. 
\newcommand{\captionseven}{Normalized CINN MAP estimated RIR (top) and posterior mean RIR and 99.7 \% confidence interval (bottom) at an extrapolated position both superimposed over the measured reference RIR}
\begin{figure*}[ht!]
\centering
\includegraphics[trim=0.1cm 0.12cm 0.1cm 0.11cm, clip,width=\linewidth]{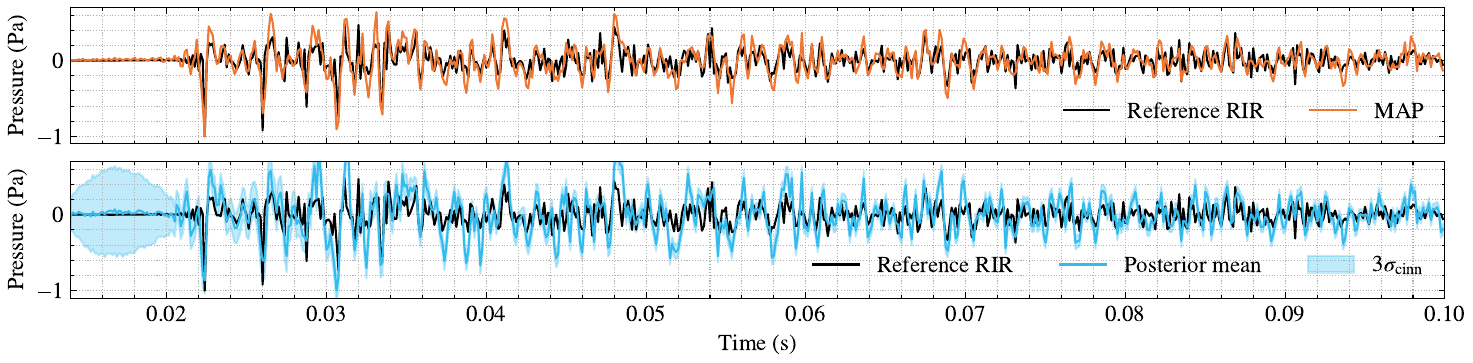}
\caption{\captionseven}
\label{fig:ExtrapRIR}
\end{figure*}

\section{Conclusions}\label{sec:conclusions}

A conditional invertible neural network, capable of learning the forward operation of plane wave propagation, has been proposed for sound field reconstruction. This approach offers several advantages over existing methodologies. Firstly, it does not necessitate exhaustive datasets for training, as the training data was simulated and obtained using random wave fields. This reduces the burden of data collection and allows for more efficient training processes.

Furthermore, the model provides interpretable uncertainty estimates, delineating mainly uncertainties apparent in the acoustic model. This allows for a better understanding of the reconstructed sound fields' reliability and the potential need for model refinement when no ground truth data is available.

Moreover, the proposed neural network can serve as an invertible prior for MAP estimation of plane wave coefficients, or be directly sampled to obtain posterior mean coefficients, conditioned on the measured pressure and parameters obtained during training. This flexibility in estimation methods broadens the versatility of the model and enables one to choose the approach that best suits their specific requirements.

In comparative evaluations and under challenging acoustic conditions, the proposed neural network outperforms the hierarchical Bayes model at high frequencies, showing improved performance in the experimental validation. Additionally, it exhibits greater consistency than the hierarchical Bayes model in terms of robustness to array decimation and variability of data. Lastly, the proposed approach is well suited for RIR reconstruction, offering either the accuracy of MAP-estimated RIRs or the rapid estimation of posterior mean RIRs.


\bibliographystyle{IEEEtran}  
\bibliography{NFFlowsPaper.bib}

\appendix
\section{Mapping variables with rational quadratic splines} \label{section:appendixA}

The splines defined by \cite{Durkan2019} employ $K$ distinct rational-quadratic functions (bins), delimited by $K+1$ coordinates known as knots $\left\{\left(z^{(k)}, y^{(k)}\right)\right\}_{k=0}^K$. These knots increase monotonically from $\left(z^{(0)}, y^{(0)}\right) = (-B,-B)$ to $\left(z^{(K)}, y^{(K)}\right) = (B, B)$. Internally the spline derivatives take on $K-1$ arbitrary positive values $\left\{\phi^{(k)}\right\}_{k=1}^{K-1}$, while the boundary derivatives are given by $\phi^{(0)}=\phi^{(K)}=1$ which act as linear `tails'.

The method constructs a monotonic, continuously-differentiable, rational-quadratic spline passing through knots with prescribed derivatives. The rational-quadratic function $h^{(k)}$ within the $k^{\text{th}}$ bin is expressed as
\begin{align}\label{eq:rational_quadratic_spline}
h(z^{(k)}) &= \frac{\alpha^{(k)}(\xi)}{\beta^{(k)}(\xi)} \nonumber \\ 
            &=  y^{(k)} + \frac{\left(y^{(k+1)}-y^{(k)}\right)\left[s^{(k)} \xi^2+\phi^{(k)} \xi(1-\xi)\right]}{s^{(k)}+\left[\phi^{(k+1)}+\phi^{(k)}-2 s^{(k)}\right] \xi(1-\xi)},
\end{align}
with $s_k=$ $\left(y^{k+1}-y^k\right) /\left(z^{k+1}-z^k\right)$ corresponding to the slopes of the lines connecting the coordinates and $\xi(z)=\left(z-z^k\right) /\left(z^{k+1}-z^k\right)$ represents the normalized distance along the $z$ axis within the $k^{\text{th}}$ bin.

The logarithm of the determinant of its Jacobian can be computed by summing the logarithm of derivatives of \eqref{eq:rational_quadratic_spline} with respect to each transformed $z$ value of the input vector. This derivative is given by
\begin{align}
\frac{\mathrm{d}h}{\mathrm{d} z} &= \frac{\mathrm{d}}{\mathrm{d} z}\left[\frac{\alpha^{(k)}(\xi)}{\beta^{(k)}(\xi)}\right] \nonumber\\
                            &= \frac{\left(s^{(k)}\right)^2\left[\phi^{(k+1)} \xi^2+2 s^{(k)} \xi(1-\xi)+\phi^{(k)}(1-\xi)^2\right]}{\left[s^{(k)}+\left[\phi^{(k+1)}+\phi^{(k)}-2 s^{(k)}\right] \xi(1-\xi)\right]^2}.
\end{align}
Analytical computation of the inverse rational-quadratic function involves inverting \eqref{eq:rational_quadratic_spline}, which in turn leads to solving the following quadratic equation
\begin{equation}
q(z) = \alpha(\xi(z)) - y \beta(\xi(z)) = a \xi(z)^2 + b \xi(z) + c = 0.
\end{equation}
Due to monotonicity, only one root is obtained. The solution is given by $\xi(z)=2 c /\left(-b-\sqrt{b^2-4 a c}\right)$, where
\begin{align}
 &a=\left(y^{(k+1)}-y^{(k)}\right)\left[s^{(k)}-\phi^{(k)}\right]  \nonumber\\
&\quad +\left(y-y^{(k)}\right)\left[\phi^{(k+1)}+\phi^{(k)}-2 s^{(k)}\right] \\
&b=\left(y^{(k+1)}-y^{(k)}\right) \phi^{(k)}\nonumber \\
&\quad -\left(y-y^{(k)}\right)\left[\phi^{(k+1)}+\phi^{(k)}-2 s^{(k)}\right] \\
&c=-s^{(k)}\left(y-y^{(k)}\right)
\end{align}
where the coefficients \( a \), \( b \), and \( c \) depend on the target output \( y \). These expressions allow us to determine \( \xi(z) \), subsequently enabling the computation of the inverse transformation of $h$.
\ifdefined\appendcaptions
\section*{Figure captions}
\renewcommand{\labelenumi}{Figure \arabic{enumi}:}
\begin{enumerate}
\setlength{\itemindent}{5ex}
\item{\captionzero}
\item{\captionzeroone}
\item{\captionzerotwo}
\item{\captionone}
\item{\captiontwo}
\item{\captionthree}
\item{\captionfour}
\item{\captionfive}
\item{\captionsix}
\item{\captionseven}
\item{\captioneight}
\item{\captionnine}
\item{\captionten}
\item{\captioneleven}
\item{\captiontwelve}
\end{enumerate}
\fi
\end{document}